\documentclass[a4paper,11pt]{article}
\pdfoutput=1 

\usepackage{jheppub} 

\usepackage[T1]{fontenc} 

\usepackage{mathtools}
\usepackage{bm}

\newcommand{\bref}[1]{(\ref{#1})}
\newcommand{\dl}{\partial}
 
\newcommand{\ii}{\mathrm{i}}
\newcommand{\AD}{\dot{A}}
\newcommand{\BD}{\dot{B}}
\newcommand{\CD}{\dot{C}}
\newcommand{\DD}{\dot{D}}
\newcommand{\fa}{\mathfrak{a}}
\newcommand{\fb}{\mathfrak{b}}
\newcommand{\fc}{\mathfrak{c}}
\newcommand{\fd}{\mathfrak{d}}
\newcommand{\fr}{\mathfrak{r}}
\newcommand{\ehat}{\hat{e}}
\newcommand{\hGamma}{\hat{\Gamma}}
\newcommand{\tGamma}{\tilde{\Gamma}}
\DeclareMathOperator{\tr}{tr}
\newcommand{\dd}{\mathrm d}

\newcommand{\eps}{\epsilon}
\newcommand{\lamphi}{\lambda_{\phi}}

\newcommand{\tSigma}{\tilde{\Sigma}}
\newcommand{\Boxeta}{\Box_{\eta}}

\title{\boldmath Spinors, Doppelg\"angers and the Non-Perturbative Classical Double Copy}

\author[a,1]{Kymani Armstrong-Williams\note{Corresponding author.}}

\affiliation[a]{Centre for Theoretical Physics and Astronomy, School of Physical and Chemical Sciences,
Queen Mary University of London, 327 Mile End Road, London E1 4NS, UK
}

\emailAdd{k.t.k.armstrong-williams@qmul.ac.uk}

\abstract{We develop a geometric and spinorial formulation of the non-perturbative classical double copy between the Corrigan--Fairlie--'t Hooft--Wilczek sector of complexified SU(2) Yang--Mills theory and four-dimensional conformally flat gravity. We identify the gauge potential with the positive-chiral spin connection of the conformally flat metric, establishing an exact relation between the Yang--Mills field strength and the chiral projection of the Riemann tensor. This provides a four-dimensional realisation of the doppelg\"anger viewpoint, in which gauge and gravitational fields arise from different interpretations of common connection data. Within this sector, the sourceless Yang--Mills equations require constant scalar curvature, explaining the shared nonlinear scalar equation and extending the gravitational correspondence from traceless matter to matter with constant energy--momentum trace. By representing adjoint colour indices as symmetric pairs of spinor indices, we derive exact relations mapping the gauge field-strength spinors to the scalar-curvature and trace-free Ricci components of the gravitational curvature. These relations retain nontrivial curvature information despite the vanishing Weyl tensor. We illustrate the construction using anti-de Sitter and de Sitter spacetimes, together with $\mathrm{AdS}_{3}\times\mathbb{R}$ supported by a scalar field, demonstrating the extension to matter with nonzero constant trace.}

\begin{document} 
\maketitle
\flushbottom

\section{Introduction}
\label{sec:intro}
A full realisation of the exact non-linear relationship between particle physics and gravity has remained annoyingly elusive. At the level of solutions to the classical equations of motion, the classical double copy relates certain abelian and non-abelian gauge theory solutions to their counterparts in gravity \cite{CarrilloGonzalez:2026phk,White:2024pve,Monteiro:2014cda,Bahjat-Abbas:2017htu,Luna:2015paa,Carrillo-Gonzalez:2017iyj,Luna:2016due,Armstrong-Williams:2022apo}.\footnote{see also refs. \cite{Ilderton:2024oly,Anastasiou:2014qba,Anastasiou:2018rdx,Luna:2020adi,Borsten:2021zir,Godazgar:2022gfw,Liang:2023zxo} for a discussion on the Convolutional Double Copy.}
First observed for $D \ge 4$ Kerr-Schild vacuum solutions in general relativity, the correspondence quickly was extended to all Petrov type $D$ and $N$ solutions. In addition, these results were modified to make extensions to thermodynamic quantities, $\mathcal{N}=0$ supergravity, non-vacuum solutions in Einstein-Maxwell gravity and certain Petrov type $\mathrm{I}$ solutions \cite{Godazgar:2020zbv,White:2020sfn,Chacon:2021wbr,Luna:2018dpt,Alawadhi:2020jrv,Easson:2021asd,Luna:2022dxo,Chacon:2020fmr,Keeler:2024bdt,Keeler:2020rcv,Armstrong-Williams:2023ssz,Zhao:2024ljb,Zhao:2024wtn,Kent:2024mow,Caceres:2025eky,Kent:2025pvu,Moynihan:2025vcs,CarrilloGonzalez:2022ggn,Emond:2022uaf,Armstrong-Williams:2024bog,Wang:2026ojn,LaCruz:2026lvf,Easson:2026vep,Bagchi:2026pjx,Easson:2026wex,Easson:2026xod,Easson:2023dbk,Easson:2020esh}.
\\
\\
Previously, the main stumbling block for this program of research was the lack of true non-perturbative examples of this classical correspondence.
Progress for a non-perturbative correspondence came first in two spacetime dimensions at the level of the Lagrangian, which related together biadjoint scalar, Zakharov–Mikhailov and special Galileon theories \cite{Cheung:2022mix}.
\begin{figure}[t]
    \centering
    \includegraphics[width=\textwidth]{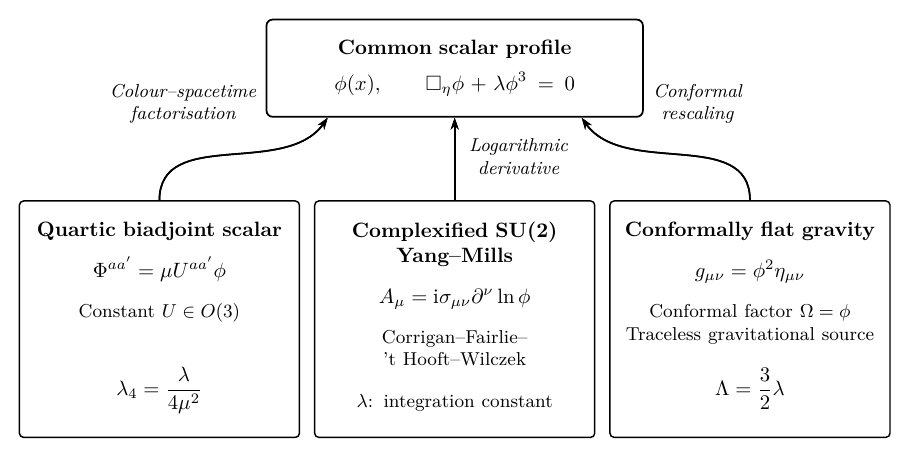}
    \caption{Common scalar description of the quartic biadjoint,
    complexified $\mathrm{SU}(2)$ Yang--Mills, and conformally flat
    gravitational ansatz sectors. With the parameter identification
    $\lambda=4\mu^{2}\lambda_{4}=2\Lambda/3$, the biadjoint and
    Yang--Mills equations, together with the trace of the Einstein
    equations for traceless matter, reduce to
    $\Box_{\eta}\phi+\lambda\phi^{3}=0$, where $\Box_{\eta}$ is the D'Alembertian for Minkowski space.
    The trace-free Einstein equations additionally determine the
    gravitational source.}
    \label{fig:common-scalar-construction}
\end{figure}
A step towards a non-perturbative classical double copy in four spacetime dimensions was taken in
Ref.~\cite{Armstrong-Williams:2026wzi}, which constructed an
exact correspondence between restricted sectors of quartic biadjoint
scalar theory (see prior and related investigations in refs \cite{Cachazo:2013iea,de2017extended,white2016exact,Armstrong-Williams:2026dmk,Armstrong-Williams:2025spu,Bahjat-Abbas:2018vgo,bastianelli2021worldline,de2023holonomic,Arkani-Hamed:2017mur,Cachazo:2013gna,Cachazo:2019ngv,Mizera:2017rqa,cachazo2020notes,frost2018biadjoint}), complexified $\mathrm{SU}(2)$ Yang--Mills theory, and
four-dimensional general relativity with conformally flat metrics and
traceless matter sources, as described by figure \ref{fig:common-scalar-construction}.
The construction employs a common scalar
profile to generate a factorised quartic biadjoint field, a
Corrigan--Fairlie--'t Hooft--Wilczek gauge potential, and the conformal
factor of the metric. With appropriate parameter identifications, the
scalar and gauge equations, together with the trace of the Einstein
equations, reduce to the equation of motion of quartic scalar theory, thus retaining nonlinear
interactions across the correspondence. Crucially, the trace-free
Einstein equations additionally determines that the gravitational source
is related to the Yang--Mills energy--momentum
tensor by conformal rescaling. Examples of gravitational solutions included in this correspondence are maximally 
symmetric spacetimes and Jacobi-elliptic radiation cosmologies, while
the linear branch admits the Bertotti--Robinson geometry. These
results establish a solution-generating framework beyond linearised
reductions, while remaining restricted to the chosen ansatz sectors.
The gravity solutions in figure \ref{fig:common-scalar-construction} are themselves part of the class of Petrov type O solutions.
\\
\\
Two open questions then come to mind when considering the investigation undertaken by Ref.~\cite{Armstrong-Williams:2026wzi}. 
Firstly, in line with previous classical double copy literature, is it possible to construct a spinorial counterpart for the relationship described in figure \ref{fig:common-scalar-construction}? Secondly, what is the origin of for the non-perturbative relationship between solutions? Could it be geometrical, and if so, how can understanding its origins help us extend the catalogue of solutions that can be double copied using this prescription?
\\ 
This paper explores and answers both of these questions, developing a new spinorial formulation in the absence of a non-zero Weyl tensor as well as making manifest the geometrical relation between the Yang-Mills and gravity solutions described in figure \ref{fig:common-scalar-construction}.
The spinorial relationship makes use of the colour indices on the Yang-Mills side to construct spinorial gauge quantities we can directly equate to gravitational counterparts.
\\
\\
Also, we approach the geometrical relationship through the \emph{doppelg\"anger} viewpoint
developed in Ref.~\cite{Moynihan:2025vcs}. In its three-dimensional realisation, formulated
using Chern--Simons--Witten gravity and non-abelian Chern--Simons gauge
theory, gauge and gravitational quantities are related by reinterpreting
their internal indices and generators. This suggests examining whether
a comparable identification can explain the four-dimensional correspondence
above. Rather than beginning with a prescribed squaring operation on gauge
fields, the relevant question is whether the gauge and gravitational
connections can be understood as different interpretations of the same
connection data. We show that this possibility is realised within the
conformally flat sector. Similar investigations are taken part in the context of euclidean signature and self dual configurations \cite{Landweber:2005tq,Araneda:2022iri,liu2026self,Bhoja:2024xbe,Krasnov:2012pd}. 
One main consequence of this realisation is that we can extend the gravity solutions allowed under this correspondence to include conformally flat solutions whose energy-momentum tensor has constant trace. 
\\
\\
The paper is organised as follows. Section~\ref{sec:spinors} reviews the spinorial
conventions used throughout. Section~\ref{sec:dopp} establishes the connection-level
doppelg\"anger relationship and derives the extension to constant-trace
matter. Section~\ref{sec:spinoralr} develops the corresponding non-linear spinorial
relations. Section~\ref{sec:examples} presents the anti-de Sitter, de Sitter and $\mathrm{AdS}_{3}\times\mathbb{R}$ examples. We conclude
in section~\ref{sec:discussion}.

\section{Review of Spinorial Formalism} 
\label{sec:spinors}
This paper will make use of the language of spinorial formalism for field theory to make explicit the relationship between quantities in General Relativity and complexified SU(2) Yang-Mills Theory. This formalism allows us to write all field equations in terms of two component Weyl spinors $\pi_{A}$ and conjugate Weyl Spinors $\omega_{\AD}$.\footnote{Spinorial indices in this paper will be given by (un-)dotted capital Latin letters.}
We use the mostly-plus metric convention (which is uncommon in spinorial formalism),
namely that $\eta_{\mu \nu} =\mathrm{diag}(-1,1,1,1)$ in cartesian coordinates \cite{krasnov2020formulations}. 
The raising and lowering of spinorial indices is performed by the two-dimensional Levi-Civita symbol

\begin{align}
\varepsilon_{AB} = 
\begin{pmatrix}
0 & 1 \\
-1&0
\end{pmatrix}
\end{align}

\begin{align}
    \pi_A = \epsilon_{AB}\pi^B,
    \qquad
    \pi^B = \pi_A\epsilon^{AB},
\end{align}
where the components of $\pi_{A}$ and $\pi^{A}$ are related to each other by 
\begin{align} 
\pi_0=\pi^1,\qquad \pi_1=-\pi^0,
\end{align}
with appropriate normalisations of 
\begin{align}
    \epsilon_{AB}\epsilon^{CB} = \delta_A^{C},
    \qquad
    \epsilon_{01} = 1.
\end{align}
\\
Similar definitions hold for the dotted equivalent $\epsilon^{\dot A\dot B}$.
\\
The soldering form $S^{\mu}_{A\AD}$ allows for translation of objects from tensorial to spinorial form and vice versa. We take its definition to be (in Minkowski space)

\begin{align}
S^{\mu}_{A\AD} = \frac{\ii}{\sqrt{2}} \left(I, -\sigma_{x},-\sigma_{y},-\sigma_{z} \right), \label{eq:solderingformm}
\end{align}
\\
where $I$ is the two-dimensional identity matrix and $\sigma_{i}$ are the Pauli matrices, given by 
\begin{equation}
    \sigma_x =
    \begin{pmatrix}
        0 & 1 \\
        1 & 0
    \end{pmatrix},
    \qquad
    \sigma_y =
    \begin{pmatrix}
        0 & -i \\
        i & 0
    \end{pmatrix},
    \qquad
    \sigma_z =
    \begin{pmatrix}
        1 & 0 \\
        0 & -1
    \end{pmatrix}.
\end{equation}
\color{black} 
\\
Using the above, we can derive the spinorial translation of the Minkowski metric to be 

\begin{align}
\eta_{A \AD B \BD}  = \varepsilon_{A B} \varepsilon_{\AD \BD}.
\end{align}
\\
For a generic 4-vector $V_{\mu} = \left(v_0,v_1,v_2,v_3 \right)$, its spinorial translation is given by 

\begin{align}
V_{A \AD} = S^{\mu}_{A\AD} V_{\mu}= \frac{1}{\sqrt{2}}
\begin{pmatrix}
\ii(v_{0}-v_{3}) & -\ii v_{1} - v_{2} \\ 
-\ii v_{1} + v_{2} & \ii (v_{0}+v_{3})
\end{pmatrix}, \label{eq:generic4vectortrans}
\end{align} 
\\
where the determinant of eq \eqref{eq:generic4vectortrans} is related to the norm of $V_{\mu}$ 

\begin{align}
2\mathrm{Det}\left(V_{A\AD} \right) = -v_{0}^{2} +v_{1}^{2} +v_{2}^{2} +v_{3}^{2}.
\end{align}
\\
The dot product between the vectors $V_{\mu}$ and $W^{\mu}$ is then defined  in spinorial form as 

\begin{align}
V \cdot W = V_{A\AD} W^{A\AD}. 
\end{align} 
Furthermore, the covariant derivative $\nabla_{\mu}$ of some spinor $\psi_{A}$ is given by \cite{Cardoso:2010ac}

\begin{align}
\nabla_\mu \psi_{A} = \partial_\mu \psi_{A} + \frac{\ii}{2}\omega_\mu^{\fa\fb} {{\sigma_{\fa\fb}}_{A}}^{C} \psi_{C}.  \label{eq:spinorcd}
\end{align}
\\
Here $\omega_\mu^{\fa\fb}$ is the spin connection and ${{\sigma_{\fa\fb}}_{A}}^{C}$ is the positive chiral generator of the $SO(1,3)$ algebra with components which are proportional to the SU(2) Pauli matrices $\sigma_{i}$ (momentarily suppressing spinorial indices)

\begin{align}
\sigma_{ij} = \frac{1}{2}\epsilon_{ijk}\sigma_k,\quad
\sigma_{i0} = \frac{\ii}{2}\sigma_i,\quad
\sigma_{0i} = -\frac{\ii}{2}\sigma_i,\quad
\sigma_{\mu\mu} = 0, \label{eq:lorentzgencons}
\end{align}
\\
which obeys the following commutation relation

\begin{align}
\left[\sigma_{\mu\nu},\sigma_{\rho\sigma}\right]
=
\ii \left[
\eta_{\mu\rho}\sigma_{\nu\sigma}
-\eta_{\mu\sigma}\sigma_{\nu\rho}
-\eta_{\nu\rho}\sigma_{\mu\sigma}
+\eta_{\nu\sigma}\sigma_{\mu\rho} \right ]. \label{eq:sigmacommutation}
\end{align} 
\\
The gothic indices in eq \eqref{eq:spinorcd} are local Lorentz indices, which arise from the tetrad formalism of general relativity, by constructing the metric $g_{\mu \nu}$ from a set of orthonormal frame vectors ${e^{\fa}}_{\mu}$

\begin{align}
g_{\mu \nu} = \eta_{\fa \fb} {e^{\fa}}_{\mu} {e^{\fb}}_{\nu}, \label{eq:tetrad}
\end{align}
\\
where the local Lorentz indices are raised and lowered via the local metric $\eta_{\fa \fb}$ (and its inverse $\eta^{\fa \fb}$).
The spinor affine connection coefficients $\hGamma_{\mu\,A}{}^C$ are then defined from eq \eqref{eq:spinorcd}

\begin{align}
\hGamma_{\mu\,A}{}^C = \frac{\ii}{2} \omega_\mu^{\fa\fb} {{\sigma_{\fa\fb}}_{A}}^{C}. \label{eq:hGammadef}
\end{align}
\\
The chiral Lorentz generator $\sigma_{\fa \fb}$ can also be expressed in terms of the
soldering forms as
\begin{align}
\sigma_{\fa\fb}{}_A{}^B
=
\frac{-\ii}{2}
\left(
S_{\fa}\bar{S}_{\fb}-S_{\fb}\bar{S}_{\fa}
\right)_A{}^B,
\label{eq:Sigma-def}
\end{align}
\\
with $\bar{S}_{a}$ denoting

\begin{align}
\bar{S}_{a} \equiv {S_{a}}^{A \AD} .
\end{align}
\\
For a given gauge theory with field strength tensor $F_{\mu\nu}$, its spinorial translation is given by

\begin{align}
F_{\mu \nu} \rightarrow F_{A\AD B\BD} = \phi_{AB} \varepsilon_{\AD\BD} + \widetilde{\phi}_{\AD\BD} \varepsilon_{AB}
\end{align}
\\
where $\phi_{AB}$ and  $\tilde{\phi}_{\AD \BD}$ are called the Maxwell and conjugate Maxwell spinor, respectively.
\\
For non-abelian gauge theories, we consider the colour valued field strength tensor $F^{a}_{\mu \nu}$, whose spinorial translation is given by 

\begin{align}
F^{a}_{\mu\nu} \rightarrow F^{a}_{A\AD B \BD} = \phi^{a}_{A B} \varepsilon_{\AD \BD} + \widetilde{\phi}^{a}_{\AD \BD} \varepsilon_{AB}
\label{eq:fmunuaspinor}
\end{align}
\\
In the special case of fixing the non-abelian gauge group to be $SU(2)$, the gauge index $a$ may be replaced with a symmetric pair of undotted spinorial indices $CD$ \cite{krasnov2020formulations,Cardoso:2010ac,jackiw1977spinor}, reducing eq \eqref{eq:fmunuaspinor} to be 

\begin{align}
F^{a}_{\mu\nu} \rightarrow F_{A\AD B \BD C D} = \phi_{A B C D} \varepsilon_{\AD \BD} + \widetilde{\phi}_{\AD \BD C D} \varepsilon_{AB}, 
\label{eq:fmunuaspinorsu2}
\end{align}
\\
with symmetries of 

\begin{align}
\phi_{ABCD}=\phi_{(AB)(CD)},
\qquad
\widetilde \phi_{\AD\BD CD}
=\widetilde \phi_{(\AD \BD)(CD)}.
\end{align}
\\
We can also consider the spinorial translation of the gauge covariant derivative $D_{A \AD}$, whose commutation relation (acting on some spinor $\xi^C$) is given as 

\begin{align}
\left [D_{A \AD}, D_{B \BD} \right] = \left[
\eps_{\AD\BD}\phi_{ABC}{}_D
+\eps_{AB}\widetilde{\phi}_{\AD\BD C D}{}
\right]\xi^D. \label{eq:YM-commutator}
\end{align}

\section{Conformally Flat metrics as a Doppelg\"anger of Yang-Mills Solutions} 
\label{sec:dopp}
\subsection{Solutions of Complexified SU(2) Yang-Mills theory}
In complexified SU(2) Yang-Mills theory, the Corrigan--Fairlie--'t Hooft--Wilczek solution \cite{oh1979periodic,actor1979classical,oh1985nonabelian,CORRIGAN197769} describes Minkowskian
solutions with real energy momentum, with gauge field $A_{\mu}$,

\begin{align}
A_{\mu}(x) \;=\; \ii \sigma_{\mu\nu}\, \dl^{\nu} \ln(\phi),  \label{eq:cfhsol}
\end{align}
where $\phi$ is some scalar field which satisfies 

\begin{align}
\Box \phi + \lambda \phi^{3} = 0 \label{eq:scalareom}
\end{align} 
$\lambda$ is an arbitrary constant and $\Box_{\eta}$ is d’Alembertian operator for Cartesian Minkowski space.
The colour indexed gauge field  $A^{a}_{\mu}$ is then given by

\begin{align}
A_{\mu} =g \frac{ \sigma^{a} }{2 \ii} A^{a}_{\mu}, \label{eq:colourgaugedef}
\end{align}
\\
where now the colour valued counterpart of equation \eqref{eq:cfhsol} is
\begin{align} 
A^{a}_{\mu}(x) \;=\; \frac{-2}{g} {\tSigma^{a}}_{\mu\nu}\, \dl^{\nu} \ln(\phi),
\end{align}
\\
where $\tilde{\Sigma}^{a}_{\mu \nu}$ is schematically defined from eq \eqref{eq:lorentzgencons}

\begin{align}
\sigma_{\mu \nu} = \tSigma^{a}_{\mu \nu} \sigma^{a} \label{eq:pauli-decomposition}
\end{align}
\\
To derive the form of $\tilde{\Sigma}^{a}_{\mu \nu}$, first recall that 
 the Pauli matrices are normalized according to
 \\
\begin{equation}
  \tr(\sigma^a\sigma^b)=2\delta^{ab},
  \qquad
  \tr(\sigma^a)=0.
  \label{eq:pauli-normalization}
\end{equation}
\\
Multiplying eq~\eqref{eq:pauli-decomposition} by $\sigma^b$ and taking
the trace gives
\\
\begin{align}
  \tr(\sigma^b\sigma_{\mu\nu})
  &=\tr(\sigma^b\sigma^a)\tSigma^a_{\mu\nu} \notag\\
  &=2\delta^{ba}\tSigma^a_{\mu\nu}.
\end{align}
\\
Consequently, the coefficients of the expansion are

\begin{equation}
    \tSigma^a_{\mu\nu}
    =\frac12\tr\!\left(\sigma^a\sigma_{\mu\nu}\right)
  \label{eq:sigma-coefficient-projector}
\end{equation}
\\
In the conventions used here, the nonzero components of
$\sigma_{\mu\nu}$ are

\begin{equation}
  \sigma_{ij}=\frac12\epsilon_{ijk}\sigma^k,
  \qquad
  \sigma_{0i}=-\frac{\ii}{2}\sigma^i,
  \qquad
  \sigma_{i0}=\frac{\ii}{2}\sigma^i.
  \label{eq:sigma-components}
\end{equation}
\\
We now evaluate equation~\eqref{eq:sigma-coefficient-projector} explicitly.
For two spatial indices, eqs~\eqref{eq:pauli-normalization} and
\eqref{eq:sigma-components} give

\begin{align}
  \tSigma^a_{ij}
  &=\frac12\tr\!\left(
      \sigma^a\frac12\epsilon_{ijk}\sigma^k
    \right) \notag\\
  &=\frac14\epsilon_{ijk}\tr(\sigma^a\sigma^k) \notag\\
  &=\frac14\epsilon_{ijk}(2\delta^{ak}) \notag\\
  &=\frac12\epsilon_{ija}.
  \label{eq:sigma-spatial-coefficients}
\end{align}
For one temporal and one spatial index, one similarly obtains
\begin{align}
  \tSigma^a_{0i}
  &=\frac12\tr\!\left[
      \sigma^a\left(-\frac{\ii}{2}\sigma^i\right)
    \right]
   =-\frac{\ii}{2}\delta^a_i,
  \label{eq:sigma-time-space-coefficients}\\
  \tSigma^a_{i0}
  &=\frac12\tr\!\left[
      \sigma^a\left(\frac{\ii}{2}\sigma^i\right)
    \right]
   =\frac{\ii}{2}\delta^a_i.
  \label{eq:sigma-space-time-coefficients}
\end{align}
Thus the numerical chiral bivectors are characterized by

\begin{equation}
    \tSigma^a_{ij}=\frac12\epsilon_{ija},
    \qquad
    \tSigma^a_{0i}=-\frac{\ii}{2}\delta^a_i,
    \qquad
    \tSigma^a_{i0}=\frac{\ii}{2}\delta^a_i
  \label{eq:chiral-bivector-components}
\end{equation}
\\
such that equation~\eqref{eq:chiral-bivector-components} can be written compactly
as

\begin{equation}
    \tSigma^a_{\mu\nu}
    =\frac12\epsilon^a{}_{bc}\,
      \delta^b_\mu\delta^c_\nu
     -\ii \,\delta^0_{[\mu}\delta^a_{\nu]}.
  \label{eq:chiral-bivector-compact}
\end{equation}
\\
The three bivectors are explicitly

\begin{align}
  \tSigma^1_{\mu\nu}
  &=\frac12
  \begin{pmatrix}
    0&-\ii&0&0\\
    \ii&0&0&0\\
    0&0&0&1\\
    0&0&-1&0
  \end{pmatrix},
  &
  \tSigma^2_{\mu\nu}
  &=\frac12
  \begin{pmatrix}
    0&0&-\ii&0\\
    0&0&0&-1\\
    \ii&0&0&0\\
    0&1&0&0
  \end{pmatrix},
  \notag\\[1ex]
  \tSigma^3_{\mu\nu}
  &=\frac12
  \begin{pmatrix}
    0&0&0&-\ii\\
    0&0&1&0\\
    0&-1&0&0\\
    \ii&0&0&0
  \end{pmatrix},
  \label{eq:explicit-chiral-bivectors}
\end{align}
\\
with the normalisation condition of

\begin{align}
  \tSigma^a_{\mu\nu}\tSigma^{b\,\mu\nu}
  &=2\tSigma^a_{0i}\tSigma^{b\,0i}
    +\tSigma^a_{ij}\tSigma^{b\,ij} \notag\\
  &=2\left(-\frac{\ii}{2}\delta^a_i\right)
       \left(\frac{\ii}{2}\delta^b_i\right)
    +\frac14\epsilon_{ija}\epsilon_{ijb} \notag\\
  &=\frac12\delta^{ab}+\frac12\delta^{ab} \notag\\
  &=\delta^{ab}.
  \label{eq:chiral-bivector-normalization}
\end{align}
\\
For future convenience, it is then useful to rewrite the form of equation \eqref{eq:cfhsol} by first absorbing a factor of $\ii$ into $\sigma_{\mu \nu}$ to define 

\begin{align}
\Sigma_{\mu \nu} = \ii \sigma_{\mu\nu}, \label{eq:bigsigmadef}
\end{align}
\\
which amends the commutation relations in eq \eqref{eq:sigmacommutation} for $\sigma_{\mu \nu}$  to become 

\begin{align}
[\Sigma_{ab},\Sigma_{cd}]
={}&
\eta_{bc}\Sigma_{ad}-\eta_{ac}\Sigma_{bd}
-\eta_{bd}\Sigma_{ac}+\eta_{ad}\Sigma_{bc}.
\label{eq:Lorentz-algebra}
\end{align}
\\
We also define the vector $\tGamma_{\mu}$

\begin{align}
\tGamma_\mu
&=
\Sigma_{\mu\nu}\partial^\nu\log\phi, 
\label{eq:gammatdef}
\end{align}
\\
which, when combined with eq \eqref{eq:bigsigmadef}, allows us to rewrite eq \eqref{eq:cfhsol} as

\begin{align}
A_{\mu} = \tGamma_{\mu}. \label{eq:gaugegammat}
\end{align}
\\
Reference \cite{Armstrong-Williams:2026wzi} then reinterpreted previous results in Refs.~\cite{Cervero:1978db,Cervero:1977gm,brendle2008conformal,Cadoni:2006ww,sinzinkayo1984vacuum,sinzinkayo1986new,sinzinkayo1985solutions,Guilfoyle:1998wj} to construct a non-perturbative double copy relationship between the class of complexified SU(2) Yang-Mills solutions described in eq \eqref{eq:cfhsol} and the class of conformally flat solutions to the Einstein equations with traceless energy momentum.

\subsection{Conformally Flat solutions of the Einstein Equations}
The conformally flat solutions to the
Einstein equations with traceless energy-momentum tensors all have metrics of the form

\begin{align}
g_{\mu \nu} = \phi^{2} \eta_{\mu \nu}.  \label{eq:conformalmetric}
\end{align}
\\
We can then derive the Christoffel symbols for the metric in eq \eqref{eq:conformalmetric} to be
\begin{align}
\Gamma^\rho{}_{\mu\nu}
&=\frac12g^{\rho\sigma}
\left(
\partial_\mu g_{\nu\sigma}
+\partial_\nu g_{\mu\sigma}
-\partial_\sigma g_{\mu\nu}
\right)
\nonumber\\
&=\delta^\rho_\mu v_\nu
+\delta^\rho_\nu v_\mu
-\eta_{\mu\nu}v^\rho ,
\label{eq:conformal-Christoffel}
\end{align} 
\\
where $v_{\mu}$ is defined by 

\begin{align}
v_{\mu} =\partial_\mu\log \left ( \phi \right ).
\label{eq:v-def}
\end{align}
\\
From the Christoffel symbols in eq \eqref{eq:conformal-Christoffel}, the Riemann tensor for eq \eqref{eq:conformalmetric} can be written as 

\begin{align}
{R^{\rho}}_{\mu\sigma\nu}& = \partial_{\sigma} v_{\mu}\,\delta_{\nu}^{\rho}
   + \delta_{\mu}^{\rho}\,\partial_{\sigma} v_{\nu}
   - \eta_{\mu\nu}\,\partial_{\sigma} v^{\rho} \notag \\
&- \partial_{\nu} v_{\mu}\,\delta_{\sigma}^{\rho}
   - \delta_{\mu}^{\rho}\,\partial_{\nu} v_{\sigma}
   + \eta_{\mu\sigma}\,\partial_{\nu} v^{\rho} \notag \\
&+ \delta_{\sigma}^{\rho} v_{\mu} v_{\nu}
   - \delta_{\sigma}^{\rho} v^{2}\eta_{\nu\mu}
   - \eta_{\sigma\mu} v^{\rho} v_{\nu}
   - \delta_{\nu}^{\rho} v_{\sigma} v_{\mu} \notag \\
&+ v^{2}\delta_{\nu}^{\rho}\eta_{\sigma\mu}
   + \eta_{\nu\mu} v^{\rho} v_{\sigma}, \label{eq:riemmancurvaturecon}
\end{align}
\\
with Ricci tensor $R_{\mu \nu}$

\begin{align}
R_{\mu\nu}
=-2\partial_\mu v_\nu-\eta_{\mu\nu}\partial_\rho v^\rho
+2v_\mu v_\nu-2 v^{2}\eta_{\mu\nu},
\label{eq:Ricci-v}
\end{align}
\\
and Ricci scalar $R$

\begin{align}
R=-6\phi^{-3}\Boxeta\phi, \label{eq:conformalricciscalar}
\end{align}
\\
where $\Boxeta$ is the D'Alembertian operator for Minkowski space.
\\
The Einstein field equations, with cosmological constant $\Lambda$, are

\begin{align} 
R_{\mu\nu} -\frac{1}{2} R g_{\mu \nu} + \Lambda g_{\mu \nu} = 8 \pi G_{N} T_{\mu \nu}, \label{eq:EE}
\end{align}
\\
with $G_{N}$ as the Newton's gravitational constant. Taking their trace and using the metric in eq \eqref{eq:conformalmetric} yields 

\begin{align}
\Box \phi + \frac{2}{3} \Lambda \phi^{3} = 0,  \label{eq:scalargr}
\end{align}
\\
which has the same structure as eq \eqref{eq:scalareom} upon the replacement of the coefficient of the cubic term 

\begin{align}
\frac{2}{3} \Lambda \rightarrow \lambda. 
\end{align}
The condition that the energy-momentum be traceless was first noted in Refs.~\cite{sinzinkayo1985solutions,sinzinkayo1986new,Cervero:1978db} and reiterated in Ref.~ \cite{Armstrong-Williams:2026wzi} as a necessary condition to allow the direct mapping from the SU(2) Yang Mills ansatz in eq \eqref{eq:cfhsol} to the gravitational solutions in eq \eqref{eq:scalargr}.
\\
\\
In the next subsection, we'll utilise the tetrad formalism to construct a tetrad for the metric in eq \eqref{eq:conformalmetric}, which will then allow us to reinterpret the non-perturbative relationship in figure \ref{fig:common-scalar-construction} geometrically, in the guise of a doppelg\"anger relationship. Moreover, this will enable us to extend this relationship to broad class of gravitational solutions. 

\subsection{Doppelg\"anger Relationships and Scope Extension}
We can choose to construct a tetrad for eq \eqref{eq:conformalmetric}, namely from eq \eqref{eq:tetrad} 

\begin{align}
{e_{\mu}}^{\fa} = \phi \delta^{\fa}_{\mu},  
\qquad 
{e^{\mu}}_{\fa} = \phi^{-1} \delta_{\fa}^{\mu},  \label{eq:tetradveccon}
\end{align}
\\
which can then be used to write down an orthonormal coframe
\begin{equation}
e^\fa=\phi\,\dd x^\fa,
\qquad
g=\eta_{\fa\fb}e^\fa e^\fb .
\label{eq:conformal-coframe}
\end{equation}
\\
Recalling that the first Cartan structural equation is given by 

\begin{align}
\dd e^\fa+\omega^\fa{}_\fb\wedge e^\fb=0,
\label{eq:Cartan-I}
\end{align}
\\
the exterior derivative of eq \eqref{eq:conformal-coframe} is then 

\begin{align}
\dd e^\fa
&=\dd\phi\wedge\dd x^\fa
=\dd\log\phi\wedge e^\fa .
\label{eq:de}
\end{align}
\\
Combining equations \eqref{eq:Cartan-I} and \eqref{eq:de} then allows us to deduce that the spin
connection for the metric for conformally flat solutions to the Einstein equations in eq \eqref{eq:conformalmetric} 
is therefore given as 

\begin{align}
\omega^{\fa\fb}
=
\partial^\fb\log\phi\,\dd x^\fa
-\partial^\fa\log\phi\,\dd x^\fb,
\label{eq:omega-form}
\end{align}
\\
whose components are

\begin{align}
\omega_\mu{}^{\fa\fb}
&=\delta_\mu^\fa\partial^\fb\log\phi
-\delta_\mu^\fb\partial^\fa\log\phi \notag \\
&=2\delta_\mu^{[\fa}\partial^{\fb]}\log\phi .
\label{eq:omega-components}
\end{align}
\\
Using equation \eqref{eq:omega-components}, we can calculate the spinor affine connection coefficients via direct substitution into eq \eqref{eq:hGammadef} 

\begin{align}
\hGamma_{\mu\,A}{}^B
&=\frac{1}{2}
\left(
\delta_\mu^\fa\partial^\fb\log\phi
-\delta_\mu^\fb\partial^\fa\log\phi
\right)\Sigma_{\fa\fb}{}_A{}^B.
\notag\\
&=
\delta_\mu^\fa \left(\Sigma_{\fa\fb}{}_{A}{}^{B}\partial^\fb\log\phi \right). \label{eq:hgammagrav}
\end{align}
\\
We can switch between local Lorentz indices and coordinate indices by making use of the frame vectors we defined in eq \eqref{eq:tetradveccon}.
Considering just the RHS of eq \eqref{eq:hgammagrav}, we see that (again suppressing spinorial indices)

\begin{align}
&\delta_\mu^\fa e_{\nu}^{\fb} e^{\nu}_{\fc} e^{\rho}_{\fd} e_{\rho}^{\fr} \left(\Sigma_{\fa\fb}\eta^{\fc \fd}\partial_\fr\log\phi \right) \notag \\
&= \delta_\mu^\fa  (\phi \delta_{\nu}^{\fb} ) (\phi^{-1} \delta^{\nu}_{\fc} ) ( \phi^{-1}\delta^{\rho}_{\fd} ) (\phi \delta_{\rho}^{\fr} )\left(\Sigma_{\fa\fb}\eta^{\fc \fd}\partial_\fr\log\phi \right)  \notag
\\ 
&= \delta_\mu^\fa  (\delta_{\nu}^{\fb} ) (\delta^{\nu}_{\fc} ) (\delta^{\rho}_{\fd} ) (\delta_{\rho}^{\fr} )\left(\Sigma_{\fa\fb}\eta^{\fc \fd}\partial_\fr\log\phi \right). \label{eq:RHSgammah1}
\end{align}
\\
It is also useful to recall that the frame vectors $\hat{e}^{\fa}_{\mu}$ for Minkowski space in cartesian coordinates are

\begin{align}
{\hat{e}_{\mu}}^{\fa} = \delta^{\fa}_{\mu},  
\qquad 
{\hat{e}^{\mu}}_{\fa} =  \delta_{\fa}^{\mu}.  \label{eq:tetradvecmin}
\end{align}
\\
Comparing eq \eqref{eq:tetradvecmin} with eq \eqref{eq:RHSgammah1}, we may deduce that

\begin{align}
 \delta_\mu^\fa  (\delta_{\nu}^{\fb} ) (\delta^{\nu}_{\fc} ) (\delta^{\rho}_{\fd} ) (\delta_{\rho}^{\fr} )\left(\Sigma_{\fa\fb}\eta^{\fc \fd}\partial_\fr\log\phi \right) \equiv \ehat_\mu^\fa  (\ehat_{\nu}^{\fb} ) (\ehat^{\nu}_{\fc} ) (\ehat^{\rho}_{\fd} ) (\ehat_{\rho}^{\fr} )\left(\Sigma_{\fa\fb}\eta^{\fc \fd}\partial_\fr\log\phi \right),  \label{eq:RHSgammah2}
\end{align}
\\
i.e the set of conformally flat solutions to the Einstein equations admits a spin connection for which the coordinate indices are in Cartesian Minkowskian coordinates.
This is made manifest when fully evaluating eq \eqref{eq:RHSgammah2} 

\begin{align}
\hGamma_{\mu} &=  \ehat_\mu^\fa  (\ehat_{\nu}^{\fb} ) (\ehat^{\nu}_{\fc} ) (\ehat^{\rho}_{\fd} ) (\ehat_{\rho}^{\fr} )\left(\Sigma_{\fa\fb}\eta^{\fc \fd}\partial_\fr\log\phi \right) \notag \\
&=  \Sigma_{\mu\nu}\eta^{\nu \rho}\partial_\rho \log\phi \notag \\
 &= \Sigma_{\mu\nu}\partial^\nu \log\phi. 
  \label{eq:RHSgammah3}
\end{align}
\\
By directly comparing the result in equation \eqref{eq:RHSgammah3} with the gauge field for the Corrigan--Fairlie--'t Hooft--Wilczek solution in eq \eqref{eq:gaugegammat},
one can deduce that 

\begin{align} 
A_{\mu} = \tGamma_{\mu} \equiv \hGamma_{\mu} \label{eq:gaugefieldidentity}
\end{align}
\\
namely, that the gauge field for the Corrigan--Fairlie--'t Hooft--Wilczek solution is geometrically identical to the spinor affine connection coefficients, and thus to the spin connection itself. More significantly, this allows us to identity that the gauge covariant derivative $D_{\mu}$ with the gravitational covariant derivative $\nabla_{\mu}$, namely (provided that both the Yang-Mills solutions and gravitational metric are in the same frame)

\begin{align}
D_{\mu} \equiv \nabla_{\mu}, \label{eq:covdequiv}
\end{align}
\\
for this specific set of solutions. 
Using the results above, we can broaden this geometrical mapping between quantities in SU(2) Yang-Mills theory and gravity to apply to the field strength and curvature tensors. 
Recall that the matrix-valued field strength for Yang-Mills theory in Minkowski space is given by 

\begin{align}
F_{\mu\nu}
&=
\partial_\mu A_\nu-\partial_\nu A_\mu
+[A_\mu,A_\nu] \label{eq:YMFSDEF}
\end{align}
\\
and using the definition of $v_{\mu}$ in eq \eqref{eq:v-def}, we can amend the definition of $\tGamma_{\mu}$ to be 

\begin{align}
\tGamma_{\mu}=\Sigma_{\mu\rho}v^\rho,
\end{align}
\\
which when applied to the field strength tensor in eq \eqref{eq:YMFSDEF}, yields

\begin{align}
F_{\mu\nu}
&=
\partial_\mu\Gamma_\nu-\partial_\nu\Gamma_\mu
+[\Gamma_\mu,\Gamma_\nu]
\nonumber\\
&=
\Sigma_{\nu\rho}\partial_\mu v^\rho
-\Sigma_{\mu\rho}\partial_\nu v^\rho
+[\Sigma_{\mu\rho},\Sigma_{\nu\sigma}]v^\rho v^\sigma .
\label{eq:F-direct-one}
\end{align}
\\
Applying the commutator for $\Sigma_{\mu \nu}$ in eq \eqref{eq:Lorentz-algebra} to eq \eqref{eq:F-direct-one} yields 

\begin{align}
F_{\mu\nu}
={}&
\Sigma_{\nu\rho}\partial_\mu v^\rho
-\Sigma_{\mu\rho}\partial_\nu v^\rho
+v_\nu\Sigma_{\mu\rho}v^\rho
-v_\mu\Sigma_{\nu\rho}v^\rho
-v^2\Sigma_{\mu\nu},
\label{eq:F-direct-final}
\end{align} 
\\
where direct comparison to the Riemann curvature tensor in eq \eqref{eq:riemmancurvaturecon} calculated for the choice of metric in eq \eqref{eq:conformalmetric} yields 

\begin{align}
\boxed{
F_{\mu\nu}
=\frac{1}{2}R_{\mu\nu \fa\fb}\Sigma^{\fa\fb}.}
\label{eq:F-Riemann}
\end{align}
\\
Equation \eqref{eq:F-Riemann} therefore
provides a four-dimensional realization of
the connection-level doppelg\"anger viewpoint developed in
reference ~\cite{Moynihan:2025vcs}, 
where in gauge and gravitational quantities
are related by reinterpreting their internal indices and generators. 
Of particular note, our eq \eqref{eq:F-Riemann} relates the gauge field strength in the \textit{Minkowski frame} directly to the Riemann curvature tensor in the frame defined by the metric in eq \eqref{eq:conformalmetric}.
The correspondence in reference \cite{Moynihan:2025vcs} was formulated using
Chern--Simons--Witten gravity in $2+1$ dimensions and non-abelian Chern–Simons gauge theory. In this work, we identify the
Corrigan--Fairlie--'t Hooft--Wilczek gauge connection with the (positive-chiral)
spin connection of the four-dimensional conformal metric in eq \eqref{eq:conformalmetric}.
\\
The relationship in equation \eqref{eq:F-Riemann}
is the chiral-spinor specialization of the covariant-derivative
identity appearing in eq.~(3.39) in \cite{Moynihan:2025vcs}.
Importantly, eq \eqref{eq:F-Riemann} is a geometric statement and does not itself require the involvement of the field
equations or traceless gravitational energy--momentum. 
The construction therefore realises the doppelg\"anger viewpoint
within a restricted four-dimensional sector, rather than establishing
an equivalence between unrestricted Yang--Mills theory and general
relativity, as was the case in 3-D for Chern--Simons--Witten gravity.
\\
Equation \eqref{eq:F-Riemann} can also be interpreted from the viewpoint of chirality. 
Specifically, we have chosen to construct the affine spin connection coefficients from a projection of the positive chiral Lorentz generator $\sigma_{\fa \fb}$ as opposed to its negative chiral counterpart $\bar{\sigma}_{\fa\fb}$.
\\
\\
From this geometrical construction, we can also derive exactly the form of equation \eqref{eq:scalargr}. 
Firstly, consider the sourceless equation of motion for Yang-Mills theory in Minkowski space 

\begin{align}
{D_{\eta}}^{\mu} F_{\mu \nu} \equiv \eta^{\mu \rho} {D_{\eta}}_{\rho}F_{\mu \nu} = 0, \label{eq:YMEOMeta}
\end{align}
with $D_{\eta}^{\mu}$ as the as the gauge covariant derivative. By eq \eqref{eq:F-Riemann} and it's implications, 
the raising and lowering of tensorial indices on the RHS of this equation is performed with the metric defined in eq \eqref{eq:conformalmetric}, whereas on the LHS of eq \eqref{eq:F-Riemann}, the indices are raised and lowered using the Minkowski metric. This subtlety means that one needs to be careful when applying \eqref{eq:F-Riemann} in cases where the lowering and raising of indices would be required. Conveniently, one such example is eq \eqref{eq:YMEOMeta}. Before we are allowed to apply eq \eqref{eq:F-Riemann} to eq \eqref{eq:YMEOMeta}, we must first write eq \eqref{eq:YMEOMeta} in the frame of the metric eq \eqref{eq:conformalmetric}, which gives 

\begin{align}
{D_{g}}^{\mu} F_{\mu \nu} &\equiv g^{\mu \rho} {D_{g}}_{\rho}F_{\mu \nu}, \notag \\
&\equiv \phi^{-2} \eta^{\mu\rho} {D_{g}}_{\rho}F_{\mu \nu} = 0
\label{eq:YMEOMg}
\end{align}
where ${D_{g}}_{\rho}$ is the gauge covariant derivative on curved space defined by the metric in eq \eqref{eq:conformalmetric}. 
For the choice of metric in eq \eqref{eq:conformalmetric}, eq \eqref{eq:YMEOMg} may be composed as

\begin{align}
{D_{g}}^{\mu} F_{\mu \nu} = \phi^{-2} {D_{\eta}}^{\mu}F_{\mu \nu}
\label{eq:YMEOM}
\end{align}
\\
Now that we have derived the expression in equation \eqref{eq:YMEOM}, only then are we allowed to 
apply eq \eqref{eq:covdequiv} to it to yield
\begin{align}
{ D_{\eta}}^\mu F_{\mu\nu}
= \frac{\phi^{2}}{2}
\left(\nabla^\mu R_{\mu\nu ab}\right)\Sigma^{ab} = 0.
\label{eq:YM-div-Riemann}
\end{align}
\\
It is then useful to recall that the contracted differential Bianchi identity gives

\begin{equation}
\nabla^\mu R_{\mu\nu\rho\sigma}
=\nabla_\rho R_{\sigma\nu}-\nabla_\sigma R_{\rho\nu},
\label{eq:contracted-Bianchi}
\end{equation}
\\
and that the Schouten tensor in four dimensions is defined by

\begin{align}
P_{\mu\nu}
=\frac{1}{2}
\left(
R_{\mu\nu}-\frac16Rg_{\mu\nu}
\right),
\label{eq:Schouten}
\end{align}
\\
along with the Cotton tensor 
\begin{align}
C_{\nu\rho\sigma}
=2\nabla_{[\rho}P_{\sigma]\nu}.
\end{align}
\\
Then for the choice of metric in eq \eqref{eq:conformalmetric}, the vanishing of the Weyl tensor implies \(C_{\nu\rho\sigma}=0\), which allows us to write the Ricci tensor as

\begin{align}
R_{\mu\nu}=2P_{\mu\nu}+\frac{1}{6}Rg_{\mu\nu},
\end{align}
\\
which implies the contracted differential Bianchi identity in eq \eqref{eq:contracted-Bianchi} becomes
\\
\begin{align}
\nabla_\rho R_{\sigma\nu}-\nabla_\sigma R_{\rho\nu}
&=
\frac{1}{6}
\left[
g_{\sigma\nu}\nabla_\rho R
-g_{\rho\nu}\nabla_\sigma R
\right].
\label{eq:div-Riemann-dR}
\end{align}
\\
Applying equation \eqref{eq:contracted-Bianchi} to eq \eqref{eq:YM-div-Riemann} gives 
\begin{align}
{D_{\eta}}^\mu F_{\mu\nu}
=\frac{\phi^{2}}{6}(\nabla_\rho R)\Sigma^\rho{}_\nu, 
\label{eq:YM-gradient-R}
\end{align}
\\
ultimately implying that 

\begin{align}
\nabla_\rho R = 0. \label{eq:riccigrad}
\end{align}
\\
Equation \eqref{eq:riccigrad} then implies that the Ricci scalar is constant (denoting the constant by $6\lambda_{\phi}$), which, when combined with the result for the Ricci scalar in equation \eqref{eq:conformalricciscalar} for the conformally flat metric, gives

\begin{align} 
 \Box \phi+\lamphi\phi^3=0, 
\label{eq:scalarequationgeo}
\end{align}
\\
where the identification of $\lamphi = \frac{2}{3}\Lambda$ yields equation \eqref{eq:scalargr}, thus demonstrating the geometric origin of the relationship seen in \cite{Armstrong-Williams:2026wzi,Cervero:1978db,Cervero:1977gm,brendle2008conformal,Cadoni:2006ww,sinzinkayo1984vacuum,sinzinkayo1986new,sinzinkayo1985solutions}.
\\
Interestingly, applying equation \eqref{eq:riccigrad} to the Einstein equations in eq \eqref{eq:EE} allows us to relax the condition that the trace of the energy momentum $T$ vanishes, through the constraint 
\begin{align}
\nabla_{\rho} T = 0 \label{eq:Tracevanish}
\end{align}
\\
i.e., we require that the trace of the energy momentum tensor is constant, rather than vanishing.
This new distinction then allows us to modify the form of eq \eqref{eq:scalargr} to become

\begin{align}
\Box \phi + \left( \frac{2}{3} \Lambda - \frac{4}{3} \pi G_{N} T\right) \phi^{3} = 0, \label{eq:gravexteom}
\end{align}
\\
thereby enlarging the previous correspondence between solutions to eq \eqref{eq:scalareom} in SU(2) Yang-Mills Theory and general relativity to include conformally flat solutions with an energy momentum tensor whose trace is constant.
\\
\\
In this section, we reviewed the Corrigan--Fairlie--'t Hooft--Wilczek solution of SU(2) Yang-Mills solution in Lorentzian signature \cite{oh1979periodic,actor1979classical,oh1985nonabelian,CORRIGAN197769} and the set of conformal solutions to the Einstein equations with traceless energy momentum, and the non-perturbative relationship connecting them. Using the tetrad formalism, we then derived the spin connection for solutions given by the metric in eq \eqref{eq:conformalmetric}, which we then identified with the gauge field in eq \eqref{eq:cfhsol}. 
This then allowed us to determine that the gauge covariant derivative and the covariant derivative for these solutions in both Yang-Mills and general relativity are identical. Using this relation,
we demonstrated that the non-perturbative relationship between certain solutions in SU(2) Yang-Mills theory and general relativity seen previously in \cite{Armstrong-Williams:2026wzi,Cervero:1978db,Cervero:1977gm,brendle2008conformal,Cadoni:2006ww,sinzinkayo1984vacuum,sinzinkayo1986new,sinzinkayo1985solutions} can itself be reinterpreted as doppelg\"anger relationship at the level of the field strength tensors. This result mirrored results seen in 3-D for the entirety of general relativity and non-abelian Chern–Simons gauge theory in reference \cite{Moynihan:2025vcs}.
\\
In the next section, we take the result of equation \eqref{eq:F-Riemann} and section \ref{sec:spinors} to construct a correspondence for these gravity and gauge solutions in the language of two component spinors.

\section{A Spinorial Non-Linear Relationship} 
\label{sec:spinoralr}

As previously mentioned in the introduction, the conformal metrics given in eq \bref{eq:conformalmetric} are of Petrov Type O, which means that Weyl Tensor vanishes.
As a consequence, they cannot be interpreted under the Weyl Double Copy scheme. 
\\
However, this does not rule out the possibility of other non-linear spinorial relationships. Recall that the Riemann Curvature Tensor $R_{\mu \nu \lambda \rho}$ can translated into spinorial form ${R}_{A\AD B \BD C \CD D \DD}$, which can be naturally decomposed as follows

\begin{align}
R_{A\dot{A}B\dot{B}C\dot{C}D\dot{D}}
&=
\epsilon_{\dot{A}\dot{B}}\epsilon_{\dot{C}\dot{D}}X_{ABCD}
+
\epsilon_{\dot{A}\dot{B}}\epsilon_{CD}\Phi_{AB\dot{C}\dot{D}}
\notag \\&+
\epsilon_{AB}\epsilon_{\dot{C}\dot{D}}\bar{\Phi}_{CD\dot{A}\dot{B}}
+
\epsilon_{AB}\epsilon_{CD}\bar{X}_{\dot{A}\dot{B}\dot{C}\dot{D}}.
\label{eq:RiemannSpinor}
\end{align} 
\\
where the spinorial translation of the traceless Ricci Spinor is given by (with the following symmetry property)  

\begin{align}
\Phi_{AB\dot{A}\dot{B}}
= \Phi_{(AB)(\dot{A}\dot{B})}, \label{eq:riccifreespin}
\end{align}
\\
and the spinorial field $X_{ABCD}$ is defined via (and likewise for its complex conjugate) 

\begin{align}
X_{ABCD}
=
\Psi_{ABCD}
+
\Lambda_{s}\left(
\epsilon_{AC}\epsilon_{BD}
+
\epsilon_{AD}\epsilon_{BC}
\right), \label{eq:Xspinordecomp}
\end{align} 
\\
where $\Lambda_{s}$ is the spinorial constant.
$\Psi_{ABCD}$ is the spinorial translation of the Weyl tensor $W_{\mu \nu\rho\lambda}$, which is symmetric about all of its spinorial indices

\begin{align}
\Psi_{ABCD}=\Psi_{(ABCD)}. 
\end{align}
\\
Applying the decomposition of $X_{ABCD}$ in eq \eqref{eq:Xspinordecomp} to the definition of the Riemann spinor in. eq \eqref{eq:RiemannSpinor} yields

\begin{align}
R_{A\AD B\BD C \CD D\DD}
={}&
\Psi_{ABCD}\,
\varepsilon_{\AD \BD}\,
\varepsilon_{\CD \DD}
+
\bar{\Psi}_{\AD\BD\CD\DD}\,
\varepsilon_{AB}\,
\varepsilon_{CD} \notag
\\
&+
\Phi_{AB\CD\DD}\,
\varepsilon_{\AD \BD}\,
\varepsilon_{CD}
+
\bar{\Phi}_{\AD \BD CD}\,
\varepsilon_{AB}\,
\varepsilon_{\CD \DD }\notag
\\
&+
2\Lambda_{s}\left(
\varepsilon_{AC}\varepsilon_{BD}
\varepsilon_{\AD \CD}
\varepsilon_{\BD \DD}
-
\varepsilon_{AD}\varepsilon_{BC}
\varepsilon_{\AD \DD}
\varepsilon_{\BD \CD}
\right). \label{eq:RiemannSpinor2}
\end{align}
\\
Moreover, contracting eq \eqref{eq:RiemannSpinor2} with $\epsilon^{AC}\epsilon^{\AD\CD}$ yields the spinor field $R_{B\BD D\DD}$, the spinorial translation of the Ricci tensor $R_{\mu\nu}$

\begin{align}
R_{B\BD D \DD} = - 2 \Phi_{B\BD D\DD} + 6 \Lambda_{s} \epsilon_{BD} \epsilon_{\BD \DD}, \label{eq:RicciSpinor}
\end{align}
\\
where taking it's trace (by contracting with $\varepsilon^{BD}\varepsilon^{\BD \DD}$) yields the Ricci scalar 

\begin{align} 
R = 24 \Lambda_{s} \label{eq:ricciscalarlam}
\end{align}
\\
The tensorial configuration of eq \eqref{eq:RicciSpinor} is then given by 

\begin{align}
R_{\mu\nu} = 6 \Lambda_{s} g_{\mu \nu} - 2 \Phi_{\mu \nu} \label{eq:riccitensorphi}
\end{align}
\\
where $\Phi_{\mu\nu}$ is the tensorial translation of $\Phi_{A\AD B\BD}$.
\\
Combining eqs \eqref{eq:ricciscalarlam} and \eqref{eq:riccitensorphi}  with the Einstein equations in eq \eqref{eq:EE} yields

\begin{align}
\left(3 \Lambda_{s} - \frac{\Lambda}{2} \right)g_{\mu \nu} + \Phi_{\mu \nu}    = -4\pi G_{N}T_{\mu \nu}, \label{eq:fulleinsteinphi}
\end{align}  
\\
where now 
\begin{align}
\Lambda_{s} = \frac{1}{3} \left(\frac{\Lambda}{2} - \pi G_{N} {T^{\mu}}_{\mu}\right), \label{eq:lambdasfulldef}\\ 
\Phi_{\mu \nu} = -4\pi G_{N} \left(T_{\mu \nu} - \frac{1}{4} {T^{\rho}}_{\rho} g_{\mu \nu} \right). \label{eq:phienergy}
\end{align}
\\
In addition to equations \eqref{eq:fulleinsteinphi}, \eqref{eq:lambdasfulldef} and \eqref{eq:phienergy}, the form of $\Phi_{AB\AD \BD}$ is restricted via the following constraints which follow from the Riemann tensor respecting the Bianchi identity

\begin{align}
&{\nabla^{D}}_{\CD}\Psi_{ABCD}
= {\nabla^{\DD}}_{(C}\Phi_{AB)\CD \DD}, \\
&\nabla^{B\BD}\Phi_{AB\AD\BD}
= -3\nabla_{A\AD}\Lambda_{s}.
\end{align}
\\
The spinorial translation of the covariant derivative $\nabla_{A \AD}$ obeys the following commutation relation
when acting on a spinor $\xi_{C}$

\begin{align}
[\nabla_{A\AD},\nabla_{B\BD}]\xi_C
=
\eps_{\AD \BD}X_{ABCD}\xi^D
+\eps_{AB}\Phi_{\AD \BD C D}\xi^D.
\label{eq:spin-commutator}
\end{align}
\\
Then, recalling the result of eq \eqref{eq:gaugefieldidentity}, namely that we can identify the spin connection and the SU(2) Yang-Mills gauge field with each other, we can therefore deduce an equality at the level of commutation of covariant derivatives for Yang-Mills and gravity for the specific set of solutions,

\begin{align}
\left[{D_{g}}_{A\AD}, {D_{g}}_{B\BD} \right] = \left[\nabla_{A\AD},\nabla_{B\BD} \right], \label{eq:commequal}
\end{align}
\\
where a key element of the equality of eq \eqref{eq:commequal} is that the gauge field lies in the same frame as the metric in eq \eqref{eq:conformalmetric}.
More concretely, the spinorial gauge field strength tensor $F_{A \AD B \BD C D}$ in eq \eqref{eq:fmunuaspinorsu2} for Minkowski space
is related to the spinorial gauge field strength tensor in the frame of the metric in eq \eqref{eq:conformalmetric} ${F_{g}}_{A \AD B \BD C D}$ via 

\begin{align}
{F_{g}}_{A \AD B \BD C D} = \phi^{-2} F_{A \AD B \BD C D}, \label{eq:gaugerescale}
\end{align}
where direct comparison of eq \eqref{eq:gaugerescale} to eqs \eqref{eq:spin-commutator} and \eqref{eq:YM-commutator} yields the spinorial relationship 

\begin{equation}
\boxed{
X_{ABCD} = \frac{\phi_{ABCD}}{\phi^{2}},
\qquad
\Phi_{\AD \BD C D} = \frac{\widetilde{\phi}_{\AD \BD CD}}{\phi^{2}}.}
\label{eq:exact-curvature-map}
\end{equation}
\\
Recalling that the Weyl tensor vanishes for conformally flat solutions, we can amend equation \eqref{eq:exact-curvature-map}, to yield 

\begin{align}
\phi_{ABCD}=\phi^{2} \Lambda_{s}\left(
\epsilon_{AC}\epsilon_{BD}
+
\epsilon_{AD}\epsilon_{BC}
\right),
\qquad
\widetilde{\phi}_{\AD \BD CD}= \phi^{2} \Phi_{\AD \BD C D}, 
\label{eq:exact-curvature-map-2}
\end{align}
\\
In summary, the above results show that there is indeed a spinorial double-copy-like relationship between the set of conformally flat solutions to the Einstein equation with constant-trace energy-momentum and the Corrigan--Fairlie--'t Hooft--Wilczek solution in Lorentzian signature. Where this relationship differs from the usual relationship is as follows: 

\begin{enumerate}
\item The usual spinorial or \textit{Weyl} double copy for general relativity relates the square of the Maxwell spinor for electromagnetism to the Weyl spinor \cite{Luna:2018dpt,Godazgar:2020zbv}

\begin{align}
\Psi_{ABCD} = \frac{\phi_{(AB} \phi_{CD)}}{\phi}. \label{eq:WeylDC}
\end{align}

In equation \eqref{eq:exact-curvature-map}, the spinorial relationship does not include the square of a gauge quantity. In addition, the scalar field in the denominator of eq \eqref{eq:exact-curvature-map} is squared, rather than being not squared in eq \eqref{eq:WeylDC}.

\item Since the Weyl Tensor vanishes for any metric expressible in the form of eq \eqref{eq:conformalmetric}, the spinorial relationship derived in equation \eqref{eq:exact-curvature-map} is for the mixed-index spinorial fields which make up the Riemann curvature spinor as well as the part of the Riemann curvature spinor which is proportional to the Ricci scalar.
\end{enumerate}
In the next section, we discuss specific examples of the relationship found in equation \eqref{eq:exact-curvature-map-2}. 

\section{Examples}
\label{sec:examples}
\subsection{Poincar\'e patch of anti-de Sitter space}
\label{sec:example-ads}

The line element for the Poincar\'e patch of anti-de Sitter space is given by 
(for $z>0$) 
\begin{align}
ds^{2}=\frac{L^2}{z^2}
\left(-\dd t^2+\dd x^2+\dd y^2+\dd z^2\right),
\label{eq:ads-metric}
\end{align}
\\
where comparison to eq \eqref{eq:conformalmetric} yields the form of $\phi$

\begin{align}
\phi(z)=\frac{L}{z}.
\end{align}
From the metric in \eqref{eq:ads-metric}, we then calculate the values of
\begin{align}
R=-\frac{12}{L^2},
\\
\Lambda=-\frac{3}{L^2},
\\
\Lambda_{s}=-\frac{1}{2L^2}.
\label{eq:ads-curvature-constants}
\end{align}
\\
The Yang-Mills counterpart of eq \eqref{eq:ads-metric} is then 
\begin{equation}
A_\mu
=\Sigma_{\mu \nu}\partial^\nu \log\frac{L}{z}
=-\frac1z\Sigma_{\mu 3},
\label{eq:ads-A}
\end{equation}
which physically is a static and planar gauge field, with a divergence on the $z=0$ slice.\\
Recalling that AdS is an Einstein metric (e.g $\Phi_{\mu \nu}=0$) and using the relation in eq \eqref{eq:exact-curvature-map-2} we find that 

\begin{align}
\phi_{ABCD}
=-\frac{1}{2z^2}\left [\epsilon_{AC}\epsilon_{BD}
+
\epsilon_{AD}\epsilon_{BC} \right],
\qquad
\widetilde{\phi}_{\AD\BD CD}=0.
\label{eq:ads-spinors}
\end{align}

\subsection{Flat slicing of de Sitter space}
\label{sec:example-ds}

The line element for de Sitter space with flat slicing is given by 
(for conformal time $\eta<0$) 

\begin{align}
ds^{2}= \frac{1}{H^2\eta^2}
\left(-\dd\eta^2+\dd x^2+\dd y^2+\dd z^2\right),
\label{eq:ds-metric}
\end{align}
\\
where $H^{2} = \frac{\Lambda}{3}$.
Comparing eq \eqref{eq:ds-metric} to eq \eqref{eq:conformalmetric} yields the form of $\phi$

\begin{align}
\phi(\eta)= -\frac1{H\eta}.
\end{align}
From the metric in \eqref{eq:ds-metric}, we then calculate the values of
\begin{align}
R=12H^2,
\\
\Lambda=3H^2,
\\
\Lambda_{s}=\frac{H^2}{2}.
\label{eq:ds-curvature-constants}
\end{align}
\\
The counterpart gauge field of eq \eqref{eq:ds-metric} is then 
\begin{align}
A_\mu
=\Sigma_{\mu\nu}\partial^\nu\log\phi
=\frac1\eta\Sigma_{\mu 0},
\end{align}
where the configuration is time-dependent snd spatial homogenous, which is singular at $\eta = 0$.
Again, recalling that dS is also an Einstein metric (e.g $\Phi_{\mu \nu}=0$) and using the relation in eq \eqref{eq:exact-curvature-map-2} we find that 

\begin{align}
\phi_{ABCD}
=\frac{1}{2 \eta ^{2}}\left[\epsilon_{AC}\epsilon_{BD}
+
\epsilon_{AD}\epsilon_{BC} \right],
\qquad
\widetilde{\phi}_{\AD\BD CD}=0.
\label{eq:ds-spinors}
\end{align}

\subsection{$\mathrm{AdS}_{3} \times \mathbb{R}$ }

\begin{figure}[htbp]
    \centering
    \includegraphics[width=\textwidth]
        {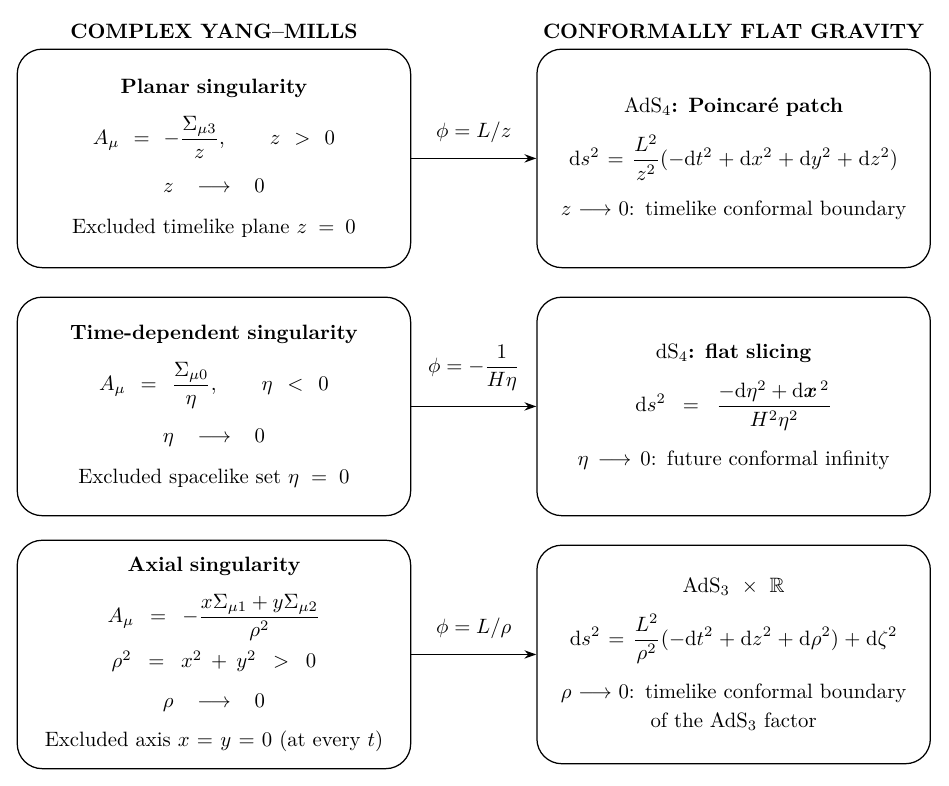}
    \caption{Gauge-field singularities and the corresponding
    gravitational boundary limits.}
    \label{fig:gauge-gravity-boundaries}
\end{figure}

The metric for Ad$\mathrm{S}_{3} \times \mathbb{R}$ in coordinates $(t,\rho,\zeta,z)$, with AdS radius $L$, is given as (for $\rho>0$) \cite{Cisterna:2017qrb,Corral:2021tww}

\begin{align}
\mathrm ds^2
=
\frac{L^2}{\rho^2}
\left(-\mathrm dt^2+\mathrm dz^2+\mathrm d\rho^2\right)
+\mathrm d\zeta^2.
\end{align}
\\
Performing a change of coordinates to cartesian coordinates via the explicit transformations

\begin{align}
x &= \rho \cos\left(\frac{\zeta}{L}\right), \notag \\
y &= \rho \sin\left(\frac{\zeta}{L}\right), \notag \\
\rho^{2} &= x^{2} + y^{2}, \label{eq:carttrans}
\end{align}
\\
yields the metric

\begin{align}
\dd s^{2} =  \frac{L^2}{x^2 + y^2}
   \left(
       -\mathrm{d}t^2
       + \mathrm{d}x^2
       + \mathrm{d}y^2
       + \mathrm{d}z^2 \right). \label{eq:ads3Rcart}
\end{align}
\\
Comparison of eq \eqref{eq:ads3Rcart} with eq \eqref{eq:conformalmetric} then determines that  $\phi$ is 

\begin{align}
\phi \equiv \phi(x,y) = \frac{L}{\sqrt{x^{2}+y^{2}}}, 
\end{align}
\\
and that the Ricci scalar is given by 

\begin{align}
R = -\frac{6}{L^{2}}.
\end{align}
The counterpart gauge field $A_{\mu}$ is then given by 
\begin{align}
A_\mu=-\frac{x\Sigma_{\mu1}+y\Sigma_{\mu2}}{x^2+y^2}, \label{eq:ads3gf}
\end{align}
which we interpret as a static cylindrically symmetric configuration which is singular at $\rho = 0$.
\\
It can then be shown that the metric in eq \eqref{eq:ads3Rcart} can be sourced by an energy momentum tensor $T_{\mu \nu}$ 

\begin{align}
 T_{\mu\nu}&=\partial_\mu\chi\,\partial_\nu\chi
 -\frac12g_{\mu\nu}g^{\alpha\beta}
 \partial_\alpha\chi\,\partial_\beta\chi, 
 \label{eq:ads3r-cartesian-matter-equations}
\end{align}
\\
Where $\chi$ is defined to be 

\begin{align}
 \chi(x,y)=\pm q L \arctan(y/x)+\chi_0, \qquad
 q^2=\frac{1}{4\pi G_NL^2}.
 \label{eq:ads3r-cartesian-matter-normalized}
\end{align}
\\
The value of $q^{2}$ is fixed by constraints imposed by the Einstein equations in eq \eqref{eq:EE}, and $\chi_{0}$ is an arbitrary constant.
The trace of the energy momentum tensor is calculated to have a constant value 

\begin{align}
T = -q^{2},
\end{align}
\\
which, importantly for the extended relationship in eqs  \eqref{eq:Tracevanish} and \eqref{eq:gravexteom}, vanishes under the application of the covariant derivative $\nabla_{\rho}$.
Also, from eqs \eqref{eq:ads3Rcart} and \eqref{eq:ads3r-cartesian-matter-equations} the value of the cosmological constant is
\begin{align}
\Lambda = - \frac{2}{L^{2}},
\end{align}
implying that 
\begin{align}
\Lambda_{s} = -\frac{1}{4L^{2}}.
\end{align}
From equation \eqref{eq:phienergy}, we can then calculate the field $\Phi_{\mu \nu}$

\begin{align}
\Phi_{\mu\nu}
=
\frac{1}{4\rho^4}
\begin{pmatrix}
-\rho^2&0&0&0\\
0&x^2-3y^2&4xy&0\\
0&4xy&y^2-3x^2&0\\
0&0&0&\rho^2
\end{pmatrix},
\label{eq:phimunu}
\end{align}
\\ 
To derive its spinorial equivalent  $\Phi_{AB \AD\BD}$, we first construct the soldering form $\hat{S}^{\mu}_{A\AD}$ for
the metric ansatz in eq \eqref{eq:conformalmetric}.
This is simply done by taking the definition of our soldering form in flat space in eq \eqref{eq:solderingformm} and contracting it with the frame vector in eq \eqref{eq:tetradveccon}

\begin{align}
\hat{S}^{\mu}_{A\AD} \equiv S^{\mathfrak{a}}_{A\AD} e_{\mathfrak{a}}{}^{\mu} = \frac{\ii \phi^{-1}}{\sqrt{2}} \left(I, -\sigma_{x},-\sigma_{y},-\sigma_{z} \right), \label{eq:solderingformc}
\end{align}
 \\
which, when applied to eq \eqref{eq:phimunu} gives 

\begin{align}
\Phi_{AB\AD\BD} &= \hat{S}^{\mu}_{A\AD} \hat{S}^{\nu}_{B\BD} \Phi_{\mu\nu} \notag \\
&=
\begin{pmatrix}
    \begin{pmatrix}
        0 & 0 \\[6pt]
        0 & -\dfrac{x-\mathrm{i}y}{2L^{2}(x+\mathrm{i}y)}
    \end{pmatrix}
    &
    \begin{pmatrix}
        0 & \dfrac{1}{4L^{2}} \\[6pt]
        \dfrac{1}{4L^{2}} & 0
    \end{pmatrix}
    \\[18pt]
    \begin{pmatrix}
        0 & \dfrac{1}{4L^{2}} \\[6pt]
        \dfrac{1}{4L^{2}} & 0
    \end{pmatrix}
    &
    \begin{pmatrix}
        -\dfrac{x+\mathrm{i}y}{2L^{2}(x-\mathrm{i}y)} & 0 \\[6pt]
        0 & 0
    \end{pmatrix}
\end{pmatrix},
\label{eq:tracefreericcispincon}
\end{align}
enabling us to construct the spinorial quantities and relationships in equation \eqref{eq:exact-curvature-map-2} 

\begin{align}
\phi_{ABCD}
&=\frac{-1}{4\rho^{2}}\left[\epsilon_{AC}\epsilon_{BD}
+
\epsilon_{AD}\epsilon_{BC} \right],
\notag \\
\widetilde{\phi}_{\AD\BD CD}&= \frac{L^{2}}{\rho^{2}}
\begin{pmatrix}
    \begin{pmatrix}
        0 & 0 \\[6pt]
        0 & -\dfrac{x+\mathrm{i}y}{2L^{2}(x-\mathrm{i}y)}
    \end{pmatrix}
    &
    \begin{pmatrix}
        0 & \dfrac{1}{4L^{2}} \\[6pt]
        \dfrac{1}{4L^{2}} & 0
    \end{pmatrix}
    \\[18pt]
    \begin{pmatrix}
        0 & \dfrac{1}{4L^{2}} \\[6pt]
        \dfrac{1}{4L^{2}} & 0
    \end{pmatrix}
    &
    \begin{pmatrix}
        -\dfrac{x-\mathrm{i}y}{2L^{2}(x+\mathrm{i}y)} & 0 \\[6pt]
        0 & 0
    \end{pmatrix}
\end{pmatrix}.
\label{eq:adsr-spinors}
\end{align}
\\
One striking observation for all the examples discussed in this section is how the singularity present in each of the Yang-Mills solutions is double copied into gravity. In particular, the gravitational counterparts of these singularities (per figure \ref{fig:gauge-gravity-boundaries}) are (conformal) boundaries of the spacetimes. 

\section{Discussions} 
\label{sec:discussion}
In this paper, we have demonstrated that there exists a geometric and spinorial interpretation
of the non-perturbative correspondence between the
Corrigan--Fairlie--'t Hooft--Wilczek sector of complexified
$\mathrm{SU}(2)$ Yang--Mills theory and conformally flat gravitational
metrics. 
\\
\\
The central observation is that the gauge potential can be
identified with the positive-chiral spinor connection of the metric
$g_{\mu\nu}=\phi^{2}\eta_{\mu\nu}$, with $\phi$ as the conformal scalar. 
This in turn allows us to construct a four-dimensional
realisation of the \emph{doppelg\"anger} viewpoint of Ref.~\cite{Moynihan:2025vcs} for our non-perturbative relationship. This allowed us geometrically to construct a doppelg\"anger relationship between the general Riemann tensor for all conformally flat solutions of the Einstein equations and the Yang-Mills field strength tensor for complexified
$\mathrm{SU}(2)$ Corrigan--Fairlie--'t Hooft--Wilczek solutions. Importantly, this gave us a geometric origin of the shared scalar
description employed in Ref.~\cite{Armstrong-Williams:2026wzi}. 
\\
\\
Imposing the Einstein equations on these results allowed us to extend the non-perturbative
relationship to conformally flat solutions with energy-momentum tensors of constant trace,
whereas the correspondence in Ref.~\cite{Armstrong-Williams:2026wzi} had previously been thought to apply only to
spacetimes with traceless energy-momentum tensors.
\\
\\
We then constructed a spinorial relationship between the particular complexified
$\mathrm{SU}(2)$ Yang--Mills theory solutions and conformally flat gravity solutions, which showed that the gravitational quantity is equal to the gauge quantity divided by the square of the scalar field which satisfies the shared scalar equation.
This was mainly achieved by exploiting the fact that any $\mathrm{SU(2)}$ adjoint colour index in the language of two -component spinors can be represented as a symmetric pair of spinorial indices. This approach differed from the standard Weyl double copy approach for a multitude of reasons, the most pressing of which is that the Weyl tensor vanishes for the set of conformally flat gravitational solutions. 
This new spinorial relationship was then exhibited for the Poincar\'e patch of $\mathrm{AdS_{4}}$ and flat slicing of $\mathrm{dS_{4}}$, before being applied to the spacetime of $\mathrm{AdS_{3}} \times \mathbb{R}$, whose energy momentum tensor has constant trace. 
\\
\\
Further work should first follow the standard pathway used in previous classical
double copy studies to determine a twistor  (or indeed ambitwistor) counterpart for this non-perturbative relationship (as seen in Refs. \cite{White:2020sfn,Chacon:2021wbr}). 
Secondly, additional work should be done to find other non-perturbative examples of a connection-level relationship beyond the one found here between conformally flat solutions and the Corrigan--Fairlie--'t Hooft--Wilczek sector of complexified $\mathrm{SU}(2)$ Yang--Mills theory.
It would be intriguing to determine which other sectors of (complexified) $\mathrm{SU(2)}$ Yang-Mills theory could exhibit similar relationships to gravity. 
It would also be interesting to explore this methodology for Yang-Mills theory valued in different gauge groups, such as $\mathrm{SU(3)}$. One stumbling block is that the trick of identifying a colour index with a symmetric pair of spinorial indices is only true for the case of picking the Yang-Mills gauge group to be $\mathrm{SU(2)}$. Any further investigations would require a new way of mapping the colour information into gravity. 
\\
Finally, we determined that the conformal boundaries of $\mathrm{AdS}_{4}$ in the Poincar\'e patch, $\mathrm{dS}_{4}$ in flat slicing and $\mathrm{AdS}_{3} \times \mathbb{R}$ can be directly mapped through this correspondence to the singularities present in their complexified SU(2) Yang-Mills gauge theory counterparts.
Further work is required to determine if this observation is related to the AdS/CFT correspondence \cite{Maldacena:1997re,Witten:1998qj,Gubser:1998bc}. 
\\
\\
In conclusion, the results presented here identify a common chiral
connection as the geometric mechanism underlying the
non-perturbative classical double copy relationship found in Ref.~\cite{Armstrong-Williams:2026wzi}. 
This identification both
explains the shared non-linear scalar equation and extends the
gravitational sector from traceless to constant-trace matter.
The newly constructed spinorial formulation of this correspondence 
illustrates how non-trivial curvature information
can participate in a double-copy-like relationship even when the
Weyl tensor vanishes. Together, these results provide a firm foundation
on which to investigate how non-abelian colour data,
gravitational curvature and matter sources can be organised within
a common geometric description.

\acknowledgments

We would like to thank Laurence Jacobs, Nathan Moynihan and Chris D. White for various discussions and feedback on the manuscript. KAW is supported by a Royal Society Career Development Fellowship.

\bibliographystyle{JHEP}
\bibliography{references.bib}


%
%
%



\end{document}